\documentclass[a4paper,11pt]{article}

\usepackage{contribution}

\newcommand{\weblink}[2][]{%
    \ifthenelse{\equal{#1}{}}%
    {\textnormal{\url{#2}}}%
    {\textnormal{\href{#2}{#1}}}%
}

\def\beq{\begin{equation}}
\def\eeq#1{\label{#1}\end{equation}}
\def\eeqn{\end{equation}}

\def\beqa{\begin{eqnarray}}
\def\eeqa#1{\label{#1}\end{eqnarray}}
\def\eeqan{\end{eqnarray}}

\let\bar=\overbar

\def\etal{{\it et al.}}

\def\Dslash{\not{\hbox{\kern-4pt $D$}}}
\def\dslash{\not{\hbox{\kern-2pt $\del$}}}

\def\msb{{\bar{\ssstyle M \kern -1pt S}}}

\newcommand{\contribution}[7][]{%
  \clearpage
  \thispagestyle{plain}
  \ifthenelse{\equal{#1}{}}
  {\hypersetup{pdftitle={#2}}}
  {\hypersetup{pdftitle={#1}}}
  \hypersetup{pdfauthor={{#3} {#4}}}
  {\centering\normalfont\LARGE\bfseries\sffamily #2 \par\nobreak}
  \lhead{}
  \chead{%
    \textit{\footnotesize XIV International Conference on Hadron Spectroscopy
      (\weblink[\textit{hadron2011}]{http://www.hadron2011.de}), 13-17 June 2011, Munich, Germany}%
  }
  \rhead{}
  \bigskip
  \begin{center}
    {#3} {#4}\ifthenelse{\equal{#6}{}}{}{\footnote{\weblink[#6]{mailto:#6}}}
    \ifthenelse{\equal{#7}{}}{}{#7} \\
    \textit{#5}
  \end{center}
  \bigskip
}

\renewcommand{\abstract}[1]{%
  \begin{center}
    \begin{minipage}{0.85\textwidth}
      \begin{footnotesize}
        #1
      \end{footnotesize}
    \end{minipage}
  \end{center}
  \bigskip
}

\begin{document}

%
%
%
%
%
{  


%

\contribution[Measurement of the double polarisation observable G]  
{Measurement of the double polarisation observable G in the reactions $\overrightarrow{\gamma}\overrightarrow{p}\rightarrow p \pi^{0}$ and $\overrightarrow{\gamma}\overrightarrow{p}\rightarrow p \eta$}  
{Marcus}{Gr\"uner}  
{Helmholtz-Institut f\"ur Strahlen- und Kernphysik, University of Bonn, Germany}  
{gruener@hiskp.uni-bonn.de}  
{for the CBELSA/TAPS collaboration}  
%

\abstract{
The excitation spectrum of the nucleon consists of several overlapping resonances. To identify these resonances and their contributions to the measured cross sections, a partial wave analysis is used. A set of at least eight, well chosen, single and double polarisation observables is needed to derive an unambiguous solution. With the Crystal Barrel/TAPS setup at ELSA, single and double polarisation observables can be measured in different reaction channels, by using the combination of a linearly or circularly polarised photon beam and a longitudinally or transversely polarised butanol frozen spin target.\\
Results of the G asymmetry measurement, using linearly polarised photons and longitudinally polarised protons, in the reactions $\overrightarrow{\gamma}\overrightarrow{p}\rightarrow p \pi^{0}$ and $\overrightarrow{\gamma}\overrightarrow{p}\rightarrow p \eta$ are presented.\\
This project is supported by the DFG (SFB/TR16).
}
%

\section{Introduction}
Photoproduction experiments can provide important contributions to the understanding of the excitation spectrum of the baryon. By using electromagnetic probes, the coupling of baryon resonances to different final states is accessable. As the total cross section is a quadratic sum of the contributing partial waves, small resonance structures are dominated by others and thus cannot be studied in the total cross section measurement. It is necessary to measure differential distributions of polarisation observables to resolve these weak resonance contributions. The single polarisation observable $\Sigma$ can be accessed by scattering a linearly polarised photon beam on an unpolarised target. Its sensitivity to interferences of partial waves provides additional informations on small resonant structures, but does not lead to an unambiguous identification of all contributing partial waves. These ambiguouities can only be disentangled by double polarisation experiments, using a polarised photon beam on a polarised target. In the case of single scalar meson production, the experiment is complete if three single and four properly chosen double polarisation observables are measured in addition to the differential cross section\cite{Tabakin}. In this case all amplitudes can be determined model-independently without descrete ambiguouities.\\
To access the double polarisation observable G, a linearly polarised photon beam was scattered on a longitudinally polarised target.

\section{Experimental Setup}
The CBELSA/TAPS experiment is located at the ELSA accelerator facility in Bonn. The electron beam of up to 3.5\,GeV can be used to produce linearly polarised photons via coherent bremsstrahlung on a diamond crystal. The energy of the beam photons is measured by a tagging spectrometer, which covers an energy range of 18\% to 85\% of the incident electron beam energy, with 96 plastic counters. A scintillating fibre hodoscope, consisting of 480 fibres, increases the energy resolution in the low energy regime to 0.2\% - 2.2\%. In the centre of the Crystal Barrel setup the frozen spin butanol target system provides longitudinally polarised protons with polarisation degrees of up to 80\%. 1230 CsI(Tl) Crystals form the main calorimeter barrel and cover a polar angle range of 30$^{\circ}$ to 156$^{\circ}$ and 2$\pi$ azimuthal angle. The main detector is equipped with a cylindrical scintillating fibre inner detector to allow charge identification. A forward detector covers the angular range down to 12$^{\circ}$. It is composed of 90 CsI(Tl) crystals in combination with 180 plastic scintillator tiles for charge identification. The MiniTAPS calorimeter closes the detector front part down to 1$^{\circ}$. Made up of 216 BaF$_{2}$ Crystals, and equipped with plastic scintillators, it provides a high granularity to account for the high count rates in the forward region.

\section{Data Analysis}
The datasets presented, were measured in 2008 and 2009, using coherent peak settings of 840\,MeV, 1032\,MeV, and 1250\,MeV to produce maximal degrees of photon polarisation of 61\%, 58\%, and 55\% respectively. An energy range of 580\,MeV to 1380\,MeV was divided into energybins of 33\,MeV width in the $\pi^{0}$ photoproduction and 100\,MeV width in the $\eta$ photoproduction case. Events with one charged and two neutral particles in the final state were selected and cuts on the coplanarity, the collinearity, the missing mass of the proton, and a time cut were applied. The target material used in the polarised target system is butanol. Besides the reactions on free protons, mesons produced off protons bound in carbon contribute to the measured count rate, which therefore has to be written as:
\begin{small}
\begin{equation}
N(\theta,\phi)=(N_{C}+N_{H})\left[1-\dfrac{(N_{C}\Sigma_{C}+N_{H}\Sigma_{H})}{(N_{C}+N_{H})} p_{\gamma}^{lin} cos(2\phi) + \dfrac{N_{H}}{(N_{C}+N_{H})} p_{z}p_{\gamma}^{lin} G sin(2\phi)\right]
\end{equation}
\end{small}
whereas $N_{C}(N_{H})$ is the pure count rate due to carbon(hydrogen), and $N(\theta,\phi)$ is the measured overall count rate. The differential count rate spectra were segmented into 12 $\theta$ bins and fitted by the following fit function:
\begin{small}
\begin{equation}
N(\theta,\phi) = A \left[1 - B\,cos(2\phi)+ C\,sin(2\phi)\right]
\end{equation}
\end{small}
Due to the carbon fraction in the target, it is not possible to determine $\Sigma_{H}$, the free proton beam asymmetry, in particular. The fitparameter B depends on a compound of the beam asymmetries of bound and free protons:
{\scriptsize$B=\dfrac{(N_{C}\Sigma_{C}+N_{H}\Sigma_{H})}{(N_{C}+N_{H})} p_{\gamma}^{lin}$}. Assuming {\scriptsize$\Sigma_{C}\approx\Sigma_{H}$}, this simplifies to {\scriptsize$B\approx\Sigma_{H} p_{\gamma}^{lin}$}, so that the results can be compared to previous measurements.\\
The double polarisation observable G can be determined by analysing the fitparameter {\scriptsize$C=\dfrac{N_{H}}{(N_{C}+N_{H})} p_{z}p_{\gamma}^{lin} G$}. G is not affected by the carbon fraction in the target explicitly, because the protons bound in carbon are not polarised. For an absolute determination of the observable, the dilution factor {\scriptsize$D=\dfrac{N_{H}}{(N_{C}+N_{H})}$} is necessary. It was determined by measuring with a hydrogen and a carbon target separately, and scaling these datasets to match the data obtained with butanol\cite{Annika}. In combination with the degrees of polarisation of the beam and the target this information was used to correct the fit results.

\section{Preliminary Results}
The measured beam asymmetry $\Sigma$ for the reaction channels $\overrightarrow{\gamma}\overrightarrow{p}\rightarrow p \pi^{0}$ and $\overrightarrow{\gamma}\overrightarrow{p}\rightarrow p \eta$ is shown in figure \ref{fig:sigma_results}. The results are in good agreement to data from the previous measurements of Graal\cite{Graal} and D. Elsner\cite{Elsner}. The data sets are compared to different PWA solutions, BnGa\cite{Anisovich}, MAID\cite{Drechsel}\cite{Chiang}, and SAID\cite{SAID}.\\
\begin{figure}[htb!]
  \begin{center}
    \includegraphics[width=0.8\textwidth]{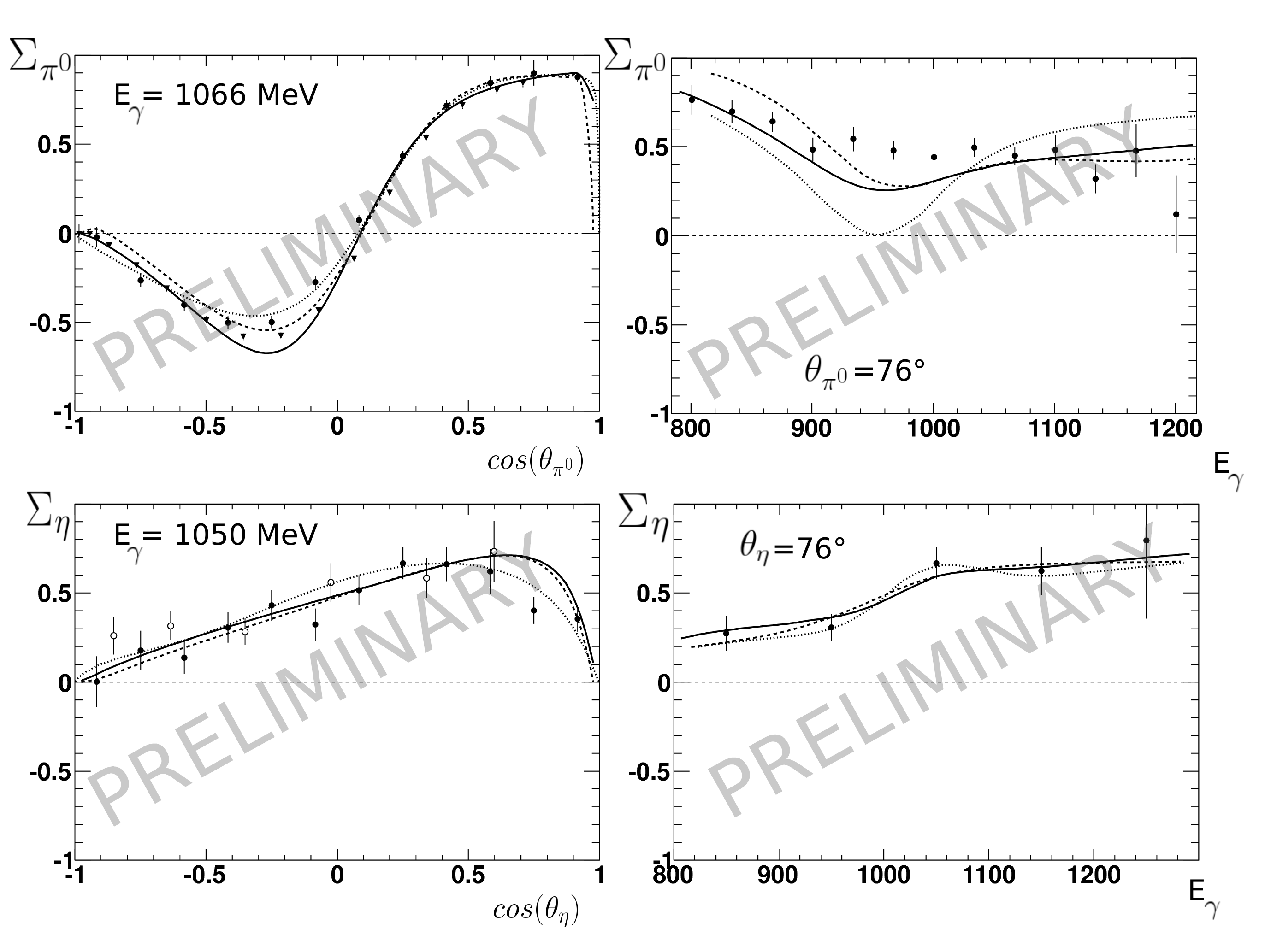}
	    \caption{\scriptsize The single polarisation observable $\Sigma$. Top: $\gamma p\rightarrow p \pi^{0}$ The preliminary results (filled circles) are compared to results of the Graal collaboration\cite{Graal} (triangles). Bottom: Preliminary results for the reaction $\gamma p\rightarrow p \eta$ (filled circles) in comparison to data measured by D.Elsner\cite{Elsner}(open circles). The lines show different PWA solutions, BnGa (solid line), SAID (dashed line), MAID (dotted line).}
    \label{fig:sigma_results}
  \end{center}
\end{figure}

\newpage
Figure \ref{fig:g_results} shows the results of the G asymmetry measurement for the reactions $\overrightarrow{\gamma}\overrightarrow{p}\rightarrow p \pi^{0}$ and $\overrightarrow{\gamma}\overrightarrow{p}\rightarrow p \eta$. The PWA solutions roughly reflect the measured data, but differences are visible. Especially in the $\eta$ photoproduction, the data cannot be discribed. These informations will provide new constraints for the prospective partial wave analysis.
\begin{figure}[htb!]
  \begin{center}
    \includegraphics[width=0.8\textwidth]{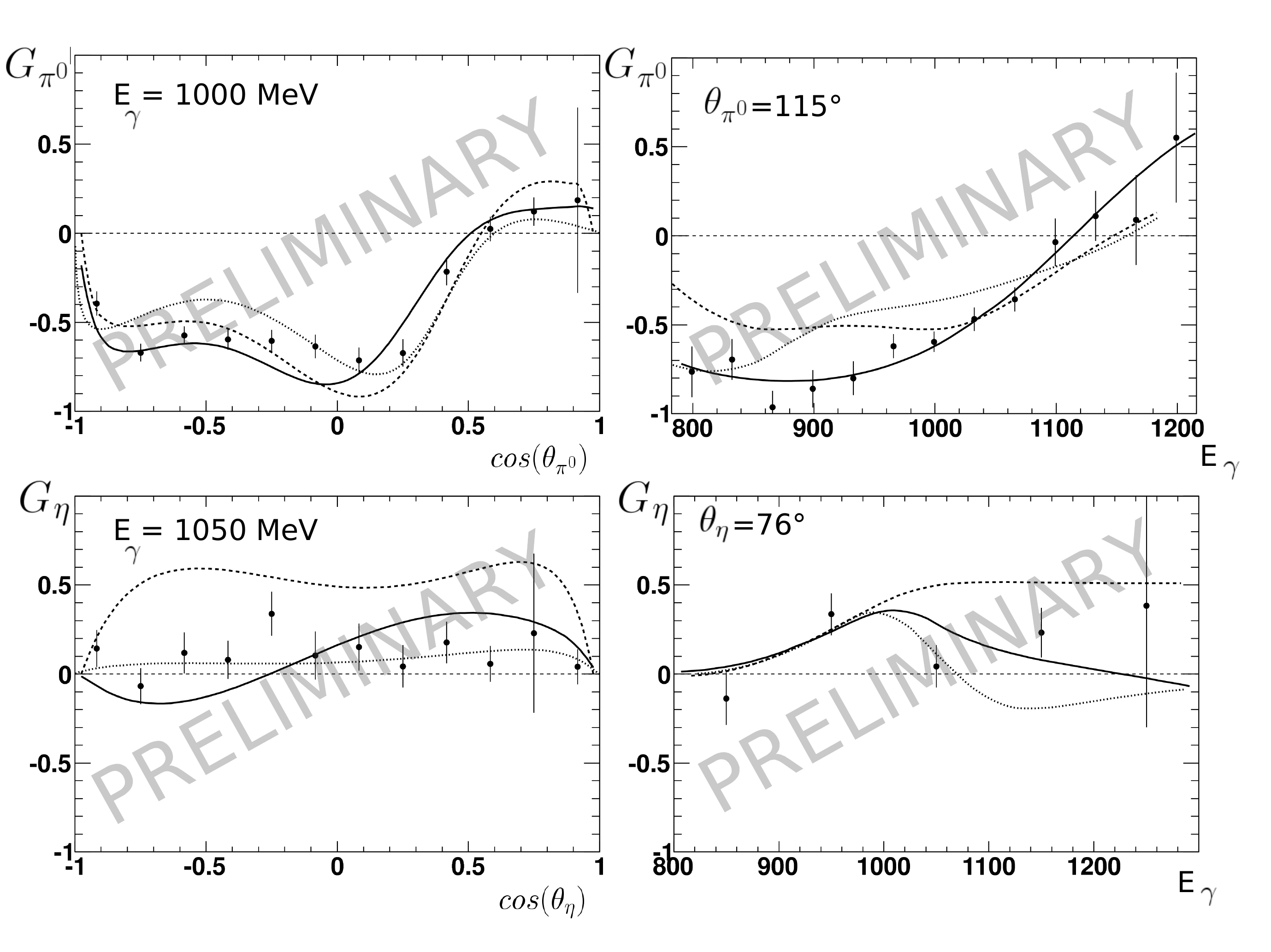}
    \caption{\scriptsize The double polarisation observable G. Top: $\gamma p\rightarrow p \pi^{0}$ bottom: $\gamma p\rightarrow p \eta$. The preliminary results (filled circles) are compared to PWA solutions, BnGa (solid line), SAID (dashed line), MAID (dotted line).}
    \label{fig:g_results}
  \end{center}
\end{figure}
\section{Summary}
{\small With the CBELSA/TAPS experiment single and double polarisation observables can be measured. The presented beam asymmetry $\Sigma$ is in good agreement with the previous measurements. The obtained results of the double polarisation observable G show discrepancies to the current PWA predictions, which will have an impact to the prospective partial wave analysis.}

%

}  


\end{document}